\documentclass[a4paper,11pt]{article}
\usepackage[utf8]{inputenc}
\usepackage{fullpage}
\usepackage{amsmath}
\usepackage{amsfonts}
\usepackage{mathtools}
\usepackage{bbm}
\usepackage{amssymb}
\usepackage{graphicx}
\usepackage{epstopdf}
\usepackage{caption}
\usepackage{color}
\usepackage{subcaption}
\usepackage{siunitx}
\usepackage{setspace}
\usepackage{psfrag}
\usepackage[square, comma, numbers, compress, sort]{natbib}
\usepackage{rotating}
\usepackage{authblk}
\usepackage{multirow}
\usepackage{multicol,float}
\usepackage[mathlines]{lineno}

\newcommand*\patchAmsMathEnvironmentForLineno[1]{%
  \expandafter\let\csname old#1\expandafter\endcsname\csname #1\endcsname
  \expandafter\let\csname oldend#1\expandafter\endcsname\csname end#1\endcsname
  \renewenvironment{#1}%
     {\linenomath\csname old#1\endcsname}%
     {\csname oldend#1\endcsname\endlinenomath}}%
\newcommand*\patchBothAmsMathEnvironmentsForLineno[1]{%
  \patchAmsMathEnvironmentForLineno{#1}%
  \patchAmsMathEnvironmentForLineno{#1*}}%
\AtBeginDocument{%
\patchBothAmsMathEnvironmentsForLineno{equation}%
\patchBothAmsMathEnvironmentsForLineno{align}%
\patchBothAmsMathEnvironmentsForLineno{flalign}%
\patchBothAmsMathEnvironmentsForLineno{alignat}%
\patchBothAmsMathEnvironmentsForLineno{gather}%
\patchBothAmsMathEnvironmentsForLineno{multline}%
}

\title{Balancing conservative and disruptive growth in the voter model}
\author[1]{Robert J. H. Ross \thanks{robert\_ross@hms.harvard.edu}}
\author[1]{W. Fontana \thanks{walter\_fontana@hms.harvard.edu}}
\affil[1]{Harvard Medical School, Department of Systems Biology, 200 Longwood Avenue, Boston, MA 02115}
\begin{document}

\date{}
\maketitle

\begin{abstract}
\noindent We are concerned with how the implementation of growth determines the expected number of state-changes in a growing self-organizing process. With this problem in mind, we examine two versions of the voter model on a one-dimensional growing lattice. Our main result asserts that the expected number of state-changes before an absorbing state is found can be controlled by balancing the conservative and disruptive forces of growth. This is because conservative growth preserves the self-organization of the voter model as it searches for an absorbing state, whereas disruptive growth undermines this self-organization. In particular, we focus on controlling the expected number of state-changes as the rate of growth tends to zero or infinity in the limit.
These results illustrate how growth can affect the costs of self-organization and so are pertinent to the physics of growing active matter.
\end{abstract}

\section{Introduction}

Many properties of growing self-organizing processes arise from an interaction between growth and activity. These include antibiotic resistance in bacterial biofilms \cite{Frost2018}, chromatophore patterning in cephalopods \cite{Reiter2018}, mammalian pigmentation patterning \cite{Mort2016}, polymer assembly \cite{Ortiz2020}, and a plethora of phenomena associated with social and technological networks \cite{Dorogovtsev2013}. To better understand how the interaction of growth and activity influences the development of such processes, we examine how the expected number of state-changes in two versions of the voter model on a one-dimensional lattice are influenced by how growth is implemented. Previous work on fully-connected growing spin models has demonstrated how different implementations of growth can interact with the functional form of the growth to determine the dynamics of the voter model. In \cite{Morris2014} it is demonstrated that depending on the combination of the implementation and functional form of growth chosen the voter model can be confined to different aspects of its phase space in the long-time limit. Similarly, other studies have been concerned with the dynamics of spin systems in which the sign of a new spin added to the system is a function of the system's current magnetisation \cite{Klymko2017,Jack2019}. 
Our work offers a different approach whereby we vary the implementation of growth via a continuous parameter, and in doing so manipulate correlations established by spin systems to control their dynamics in the large-size (long-time) limit.

As a specific motivation for our study we imagine a collection of cells that are self-organizing via local communication to achieve a desired `consensus' state.
We imagine further that this collection of cells is growing to some finite size, so that its self-organization is being repeatedly disrupted by the arrival of cells whose state must also be taken into account.  It is clear that one problem, among many, this growing self-organizing system is faced by is the following: is it more `efficient' to organize while growing, or to finish growing before self-organizing? In many instances the constraints of time and increasing combinatorial complexity suggest most complex systems would be wise to self-organize to some extent as they proceed.  Indeed, in multicellular biological systems we typically see extensive self-organization during growth. However, if \emph{how} a system grows is disruptive, that is, it undoes the previous efforts of self-organization (i.e. such as forming a pattern), may it be more efficient for the system to self-organize after it has finished growing? Continuing this theme, it is interesting to imagine to what extent growth may, in manipulating the correlations established by active matter, play a role in funneling developing systems towards desired developmental outcomes. In this work we concentrate on how growth can be used to control the expected number of state-changes in a self-organizing process. However, growth could be used to control a variety of properties.  For instance, constraining variability and adding robustness to developing systems, or allowing the same developmental processes to arrive at different developmental outcomes simply by growth being implemented in an alternative manner.

To study the effects of growth on self-organizing processes we think of these effects as belonging to two broad categories, which we refer to as `disruptive' and `conservative'. The first category contains network growth that spatially rearranges processes situated on a network.  This implementation of growth is often associated with regular networks (lattices), whereby the nodes constituting the network are constrained to a certain degree. For example, $k = 4$ in the case of a two-dimensional lattice with periodic boundaries. This means when nodes are added to the network edges are `rewired' so the network remains regular. This form of growth can have a striking effect on processes, for instance, the manipulation of spatial correlations by growth can control the outcome of population models on growing lattices \cite{Ross2016b,Ross2017b}.  The second category consists of network growth algorithms that do not rearrange processes situated on the network. This implementation of growth is often associated with complex networks, in which nodes are typically not constrained to any particular degree, and so the new node is simply `wired' to the pre-existing network. In this instance, no pre-existing edges in the network are rewired for a growth event, and so the pre-configuration of the process remains unaltered. 
This type of network growth is also associated with complex behaviors, often caused by more or less conspicuous `boundary effects', and has been studied in the context of the spread of disease \cite{Pastor2015}, game theory \cite{Poncela2008}, and the generation of traveling waves on growing networks \cite{Ross2018c}. A similar categorisation can be employed for the growth of continuous domains, however, we do not address those here.

\section{Model and results}

\subsection{Model}

Our initial lattice contains $N_{s}$ sites. 
The integer $N_{f}$ denotes the predetermined size the lattice grows to before growth is terminated.
We denote the functional form of the growth by $N(t)$.
Each site in the lattice is labelled by its position, and so the left-hand-side (LHS) boundary site is labelled $i = 1$, the site immediately next to it is labelled $i = 2$, and so on.  Each site when added to the lattice is initially either in state `0' or `1', with probability $\rho_{0}$ and $(1-\rho_{0})$, respectively.  

We study two variants of the voter model \cite{Liggett}, which we refer to as $\mathcal{C}$ (consensus) or $\mathcal{A}$ (anti-consensus). In the voter model, decision events occur at rate $P_{d}$ per site, and so the \emph{total} decision rate is $NP_{d}$. Throughout this work we set $P_{d} = 1$\footnote{As will become apparent we could have fixed our growth rate, $P_{g}$, and manipulated $P_{d}$ instead. The meaningful parameter is in fact the ratio of the total decision rate and the total growth rate.}.  
Upon a decision event a site in the lattice is chosen uniformly at random, and this site evaluates the states of its nearest neighbours before updating its own state. For a site in $\mathcal{C}$ the update probabilities are
\begin{align}
p^{\mathcal{C}}_{0\rightarrow 1} = \frac{n_{1}}{n}, \quad p^{\mathcal{C}}_{1\rightarrow 0} = \frac{n_{0}}{n},
\label{eq:c_alg}
\end{align}
where $n_{1}$ and $n_{0}$ are the number of nearest neighbors that are in state `1' or `0', respectively, and $n$ is the number of nearest neighbors (2 or 1 depending on whether the site in question is an internal site or boundary site). For a site in $\mathcal{A}$ the update probabilities are
\begin{align}
p^{\mathcal{A}}_{0\rightarrow 1} = \frac{n_{0}}{n}, \quad p^{\mathcal{A}}_{1\rightarrow 0} = \frac{n_{1}}{n}.
\label{eq:a_alg}
\end{align}
$\mathcal{C}$ and $\mathcal{A}$ will always (eventually) reach an absorbing state\footnote{Also referred to as a nonequilibrium steady-state or fixation.} for any $N_{f}$ on the one-dimensional lattice we study. In the case of $\mathcal{C}$ the absorbing states are when all sites are in state `0', or when all sites are in state `1'. For $\mathcal{A}$ the absorbing states are when every site in state `1' only has nearest neighbours in state `0', and vice versa.

\begin{figure}[H]
\centering
\vspace{-0cm}
	\includegraphics[width=0.5\textwidth]{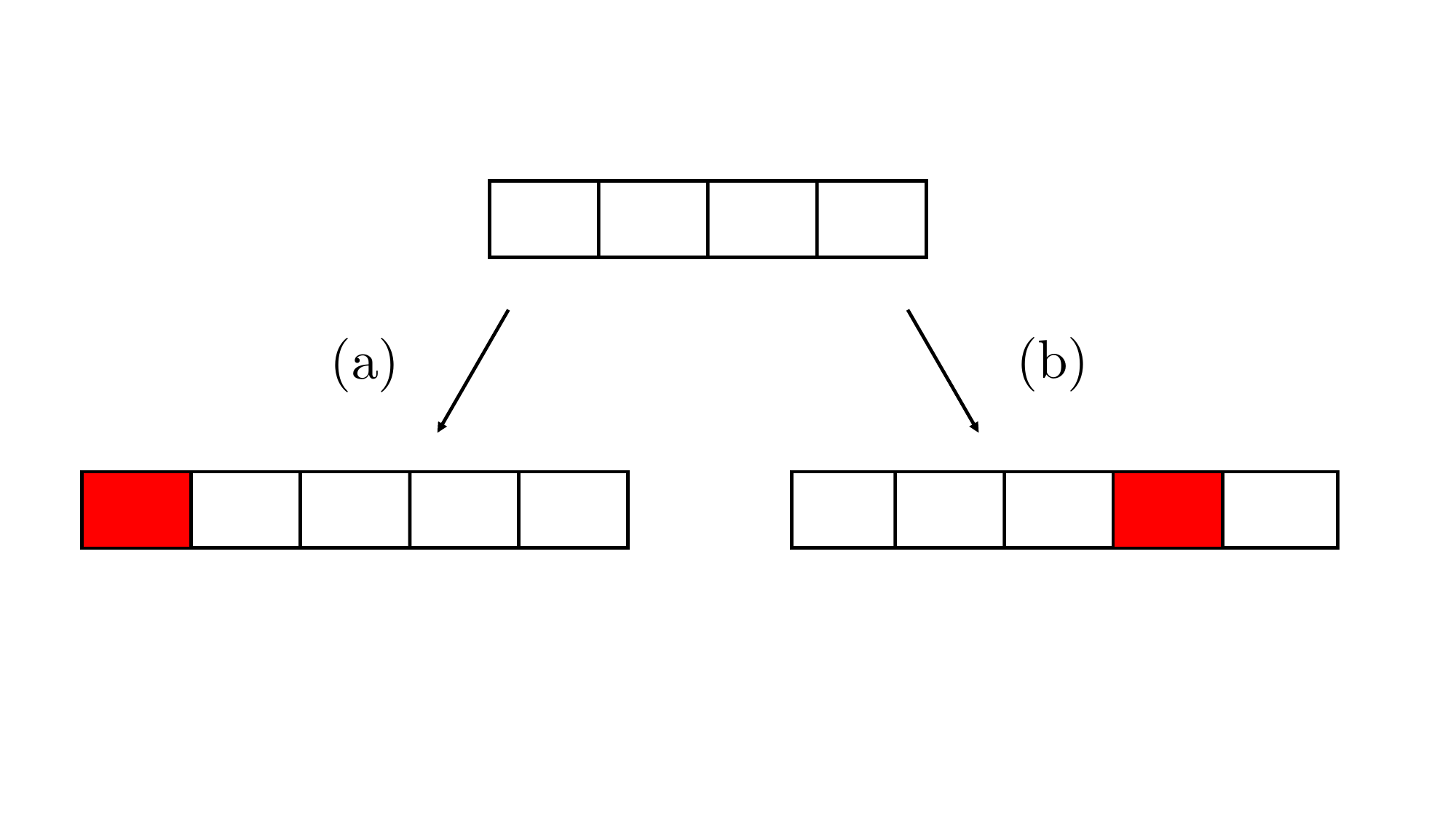}
	\vspace{-1cm}
\caption{Growth can be either conservative or disruptive. (a) This growth event, whereby a new lattice site (red) was placed on the end of the lattice, occurred with probability $\boldsymbol{g}_{4}(1)$. As no preexisting edges changed, this growth is conservative. (b) In this growth event a new lattice site was placed between sites 3 and 4, and occurred with probability $\boldsymbol{g}_{4}(4)$. This growth is disruptive as two sites that previously shared an edge do not anymore.}
\label{fig:figure1}
\end{figure}

Throughout this work we implement exponential domain growth and generalize to other functional forms of growth in section \ref{sect:PgNt}. For exponential domain growth the total domain growth rate is $NP_{g}$. We implement growth in the following manner: a vector, $\boldsymbol{g}_{N}(i)$, specifies the probability that given a growth event in a lattice of size $N$ the new site is at location $i$ in the lattice of size $N+1$.  For instance, a new lattice site is placed with probability $\boldsymbol{g}_{N}(1)$ at the LHS `end' of the lattice, and so becomes site $i = 1$ on the lattice of size $N+1$, or with probability $\boldsymbol{g}_{N}(2)$ the new site is placed in between sites 1 and 2, and so becomes site $i = 2$ on the lattice of size $N+1$, and so on. The probability of a new lattice site being placed at the right-hand-side (RHS) boundary is zero, and so $\boldsymbol{g}_{N}$ is of length $N$. In Fig. \ref{fig:figure1} we present an example of two different growth events. The general form of $\boldsymbol{g}_{N}$ can be written as
\begin{align}
\boldsymbol{\tilde g}_{N} = (\alpha_{1,N},\alpha_{2,N},\alpha_{3,N}, ... ,\alpha_{N,N}),
\label{eq:g_vector}
\end{align}
where $\sum_{i=1}^{N}\alpha_{i, N} = 1.$
The first $\boldsymbol{g}_{N}$ we study is:
\begin{align}
\boldsymbol{g}_{N}(\beta) = (1-\beta,\beta/(N-1),\beta/(N-1), ... ,\beta/(N-1)).
\label{eq:g_vector1}
\end{align}
In $\boldsymbol{g}_{N}$ the scalar $\beta$ indicates how `disruptive' growth is, with $\beta = 0$ being the most conservative form of the growth vector $\boldsymbol{g}_{N}$, and $\beta = 1$ being the most disruptive form of growth\footnote{The `most' disruptive form of growth is not well-defined here.}. Values of $\beta$ between 0 and 1 are a mixture of disruptive and conservative growth. 
As an example, to grow from a lattice of size 2 to a lattice of size 3, $\boldsymbol{g}_{N}$ is:
\begin{align}
\boldsymbol{g}_{2} = (1-\beta,\beta),
\label{eq:g_vector_example1}
\end{align}
and similarly, to grow from a lattice of size 4 to a lattice of size 5, $\boldsymbol{g}_{N}$ is
\begin{align}
\boldsymbol{g}_{4} = (1-\beta,\beta/3,\beta/3,\beta/3).
\label{eq:g_vector_example2}
\end{align}
Throughout this work we focus on the number of state-changes in models $\mathcal{C}$ and $\mathcal{A}$:
\begin{align}
\langle F^{\boldsymbol{g}(\beta),\rho_{0}} \rangle_{P_{g}} = \lim_{R\rightarrow \infty}\frac{1}{R}\sum_{r=1}^{R}F^{\boldsymbol{g}(\beta),\rho_{0}}_{P_{g},r},
\label{eq:flips}
\end{align}
where $F^{\boldsymbol{g}(\beta), \rho_{0}}_{P_{g},r}$ is the number of times any site changed (flipped) from $0 \rightarrow 1$ or $1 \rightarrow 0$ on a lattice that grew from size $N_{s}$ to size $N_{f}$ in simulation replicate $r$, with growth described by $\boldsymbol{g}$ for the specified values of $\beta$, $\rho_{0}$ and $P_{g}$. The expected number of state-changes in the model is $\langle F^{\boldsymbol{g}(\beta), \rho_{0}} \rangle_{P_{g}} + N_{f} - N_{s}$, as a growth event also results in a state-change. However, as we only compare networks grown from the same initial size, $N_{s}$, to the same size final size, $N_{f}$, we simply count `flips' instead. Our interest in counting state-changes was in part motivated by its natural association with the energetic costs of self-organization, a topic we will return to in the discussion.

Henceforth, we shall abbreviate $\langle F^{\boldsymbol{g}(\beta), \rho_{0}} \rangle_{P_{g}}$ to $\langle F^{\boldsymbol{g}(\beta)} \rangle_{P_{g}}$, and will emphasize $\rho_{0}$ when necessary. At times we will write $\langle F^{\boldsymbol{g}(\beta)}_{\mathcal{C}} \rangle_{P_{g}}$ or $\langle F^{\boldsymbol{g}(\beta)}_{\mathcal{A}} \rangle_{P_{g}}$ if a result is specific to either $\mathcal{C}$ or $\mathcal{A}$.

\subsection{Results and analysis}

Although we have presented $\mathcal{C}$ and $\mathcal{A}$ as discrete models to help build intuition, quantity \eqref{eq:flips} can be calculated exactly by imagining $\mathcal{C}$ and $\mathcal{A}$ as absorbing random walks on a directed network. The details of how to do this are standard Markov chain theory and so we describe them only briefly \cite{Howard1984}. 

The transition matrix $P$ that describes our growing voter model has $x$ transient states and 2 absorbing states for both $\mathcal{C}$ and $\mathcal{A}$:
\begin{equation}
P = 
\begin{pmatrix}
Q & R \\
\boldsymbol{0} & I_{2} \\
\end{pmatrix}
\end{equation}
where $Q$ is a $x$-by-$x$ matrix, $R$ is a nonzero $x$-by-2 matrix, $\boldsymbol{0}$ is an 2-by-$x$ zero matrix, and $I_{2}$ is the 2-by-2 identity matrix. By `state' we mean a string configuration in the voter model for a given lattice size. $Q$ describes the probability of transitioning from one transient state to another, which includes all transitions when the lattice size is below $N_{f}$, and all transitions when the lattice size is $N_{f}$ that are not into absorbing states. As such, transitions are due to either a site completing a flip or the lattice growing. $R$ describes the probabilities of transitioning from some transient state into an absorbing state at $N_{f}$, of which there are two.

From $Q$ we can obtain
\begin{align}
K = \sum_{k=0}^{\infty}Q^{k} = (I - Q)^{-1},
\label{eq:K}
\end{align}
where $K$ is referred to as the fundamental matrix. $K(p,q)$ describes the expected number of times state $q$ is visited, given the voter model started in state $p$. 
To obtain the expected number of flips we multiply the matrix $K$ by the column vector $w$ and sum over the initial states with the appropriate frequencies
\begin{align}
\langle F^{\boldsymbol{g}(\beta)} \rangle_{P_{g}} = \sum^{2^{N_{s}}}_{i=1} \gamma(i)Kw,
\label{eq:F}
\end{align}
where $\gamma$ describes the initial frequencies of each state, which depend on $\rho_{0}$, and $w$ describes the probability that given the voter model is in state $p$ the next transition is due to a site in that state successfully completing a flip, as opposed to a growth event or a decision event that does not result in a flip. For example, for state $p = (1,0,1)$ in $\mathcal{C}$ 
$$w_{p} = \frac{P_{d}}{P_{d} + P_{g}},$$
as all sites if selected would flip. Alternatively, for state $p = (1,1,0)$ in $\mathcal{C}$
$$w_{p} = \frac{P_{d}}{2(P_{d} + P_{g})}.$$

For completeness, the second moment for the number of flips is calculated via
\begin{align}
\langle (F^{\boldsymbol{g}(\beta)})^{2} \rangle_{P_{g}} = \sum^{2^{N_{s}}}_{i=1} \gamma(i)(2K_{w} - I)Kw,
\label{eq:F2}
\end{align}
where $K_{w}$ is 
\begin{align}
K_{w} = KI_{w},
\label{eq:Kw}
\end{align}
and $I_{w}$ is the diagonal matrix with $w$ on its diagonal, however, we will not make further use of Eq. \eqref{eq:F2} here.

\subsubsection{Examining different implementations of growth}

\begin{figure}[H]
\centering
    \begin{subfigure}{.3\textwidth}
	\includegraphics[width=\textwidth]{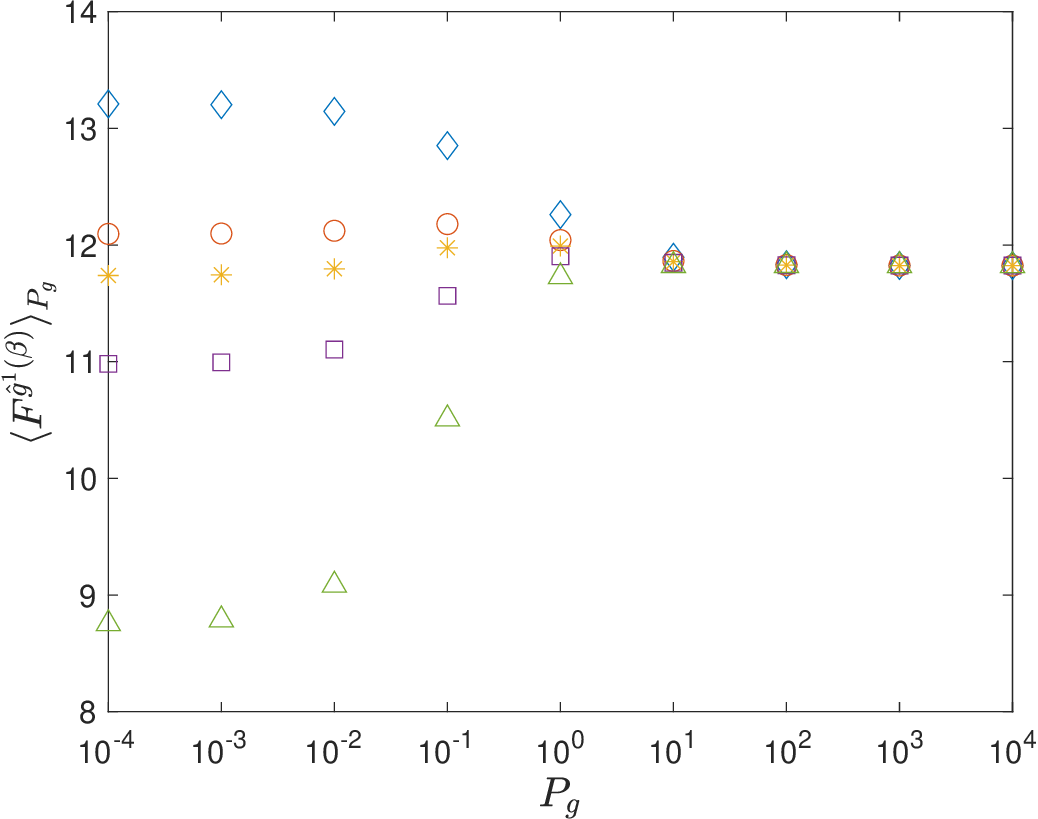}
	\subcaption{}
	\end{subfigure}
	~
	\begin{subfigure}{.3\textwidth}
	\includegraphics[width=\textwidth]{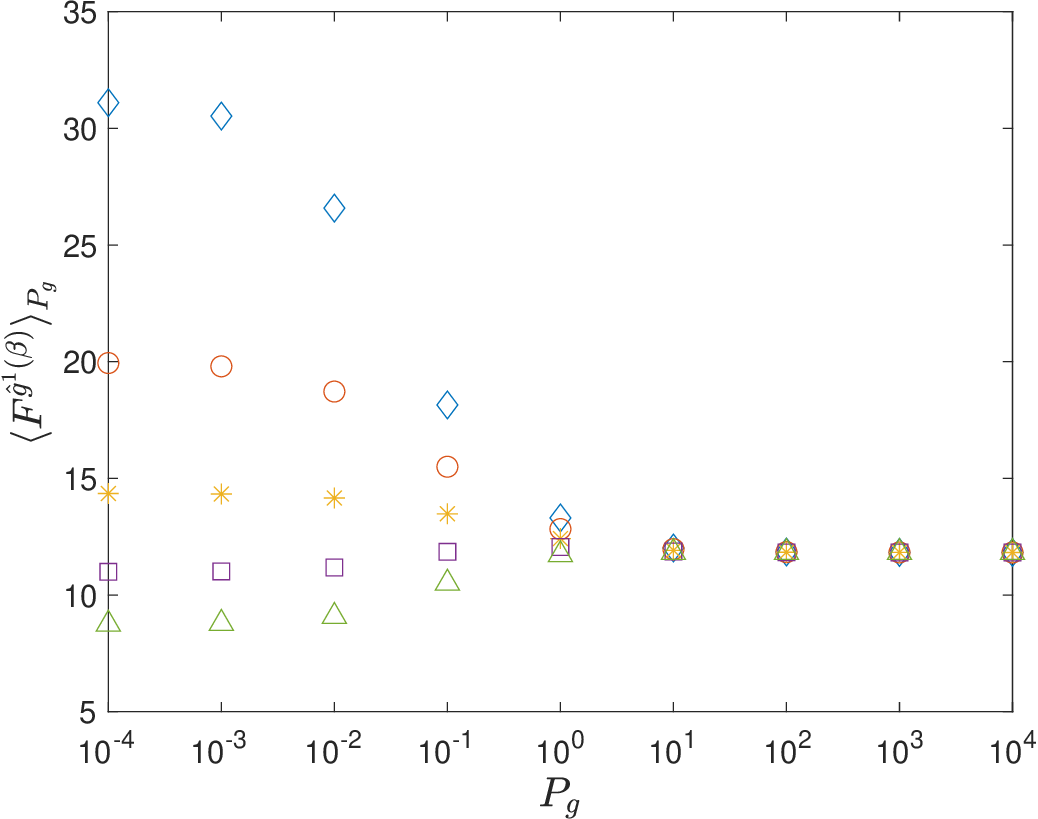}
	\subcaption{}
	\end{subfigure}
	~
    \begin{subfigure}{.3\textwidth}
	\includegraphics[width=\textwidth]{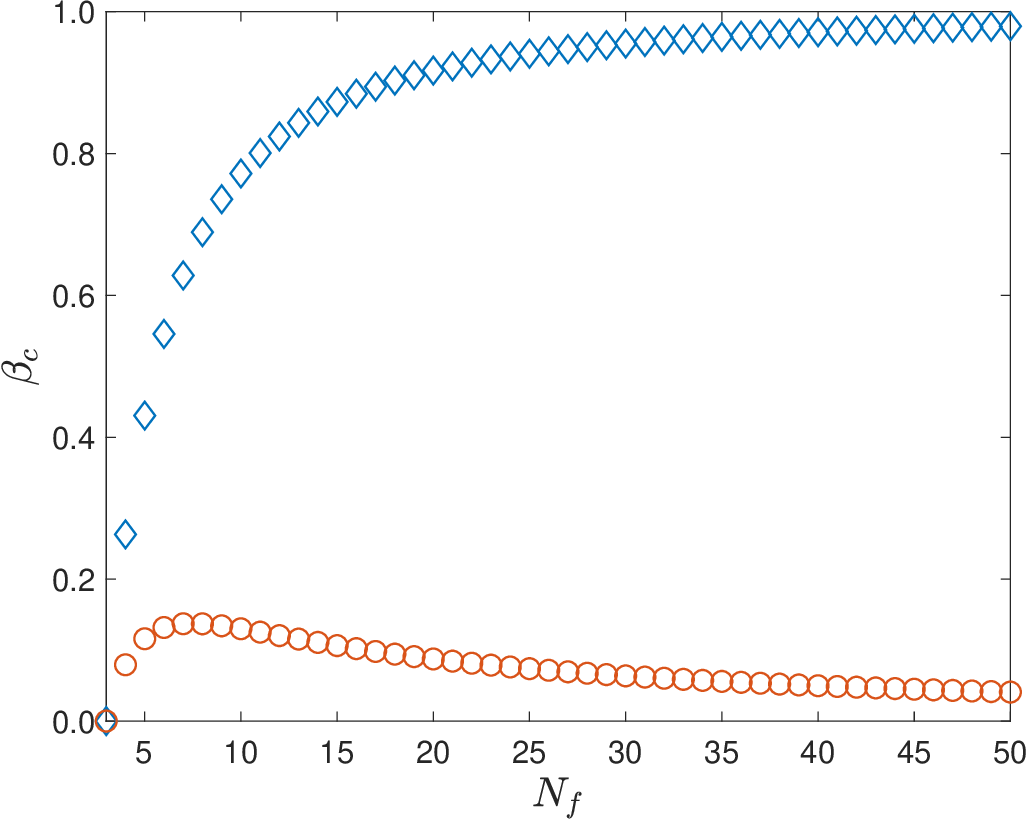}
	\subcaption{}
	\end{subfigure}
\caption{The expected number of flips in a one-dimensional lattice voter model with exponential growth from $N_{s} = 2$ to $N_{f} = 8$, with $P_{d} = 1$ and $\rho_{0} = 0.5$. (a) $\mathcal{C}$: $\beta = 1$ (blue diamonds), $\beta = 0.75$ (red circles), $\beta = 0.67$ (yellow stars), $\beta = 0.5$ (purple squares), $\beta = 0$ (green triangles). (b) $\mathcal{A}$: $\beta = 1$ (blue diamonds), $\beta = 0.5$ (red circles), $\beta = 0.25$ (yellow stars), $\beta = 0.1$ (purple squares), $\beta = 0$ (green triangles). (c) The value of $\beta_{c}$ in a one-dimensional lattice voter model grown from $N_{s} = 2$ to $N_{f}$ with $P_{d} = 1$, $\mathcal{C}$ (blue diamonds), $\mathcal{A}$ (red circles). All data points were generated using Eq. \eqref{eq:F}.}
\label{fig:figure2}
\end{figure}

In Fig. \ref{fig:figure2} (a) and (b) we display $\langle F^{\boldsymbol{g}(\beta)} \rangle_{P_{g}}$ for different values of $\beta$ and $P_{g}$ for $\mathcal{C}$ and $\mathcal{A}$, respectively.  When $\beta = 0$ the minimum value of $\langle F^{\boldsymbol{g}(\beta)} \rangle_{P_{g}}$ appears to be as $P_{g} \rightarrow 0$ in the limit for both $\mathcal{C}$ and $\mathcal{A}$. 
Conversely, when $\beta = 1$ the maximum value of $\langle F^{\boldsymbol{g}(\beta)} \rangle_{P_{g}}$ appears to be as $P_{g} \rightarrow 0$ in the limit for both $\mathcal{C}$ and $\mathcal{A}$. 
As $P_{g}$ increases the data points associated with different values of $\beta$ `coalesce'. 
The limit $P_{g} \rightarrow \infty$ is to be thought of as approaching $N_{s} = N_{f}$, and in this instance the value of $\beta$ becomes irrelevant. In the limit $P_{g} \rightarrow \infty$ it is the case that
$$\lim_{P_{g} \rightarrow \infty}\langle F^{\boldsymbol{g}(\beta), \rho_{0} = 0.5}_{\mathcal{C}} \rangle_{P_{g}} = \lim_{P_{g} \rightarrow \infty}\langle F^{\boldsymbol{g}(\beta), \rho_{0} = 0.5}_{\mathcal{A}} \rangle_{P_{g}}, \ \forall \ N_{f}.$$
From Fig. \ref{fig:figure2} (a) and (b) it can be seen that for both $\mathcal{C}$ and $\mathcal{A}$ there exists a value of $\beta$, which we refer to as $\beta_{c}$, whereby
\begin{align}
\lim_{P_{g} \rightarrow 0}\langle F^{\boldsymbol{g}(\beta_{c})} \rangle_{P_{g}} = \lim_{P_{g} \rightarrow \infty}\langle F^{\boldsymbol{g}(\beta)} \rangle_{P_{g}},
\label{eq:a1}
\end{align}
which from now on we write as
\begin{align}
\langle F^{\boldsymbol{g}(\beta_{c})} \rangle_{P_{g} \rightarrow 0} = \langle F^{\boldsymbol{g}(\beta)} \rangle_{P_{g} \rightarrow \infty}.
\label{eq:a2}
\end{align}
Equation \eqref{eq:a2} describes that when $\beta = \beta_{c}$, the expected number of flips in $\mathcal{C}$ or $\mathcal{A}$ before the absorbing state is found at $N_{f}$ is the same whether the growth rate tends to zero or infinity in the limit. This is because when $\beta = \beta_{c}$ the disruptive and conservative forces of growth are balanced in the necessary way as $P_{g} \rightarrow 0$ in the limit. 

We now concern ourselves with the following two questions: 

i) What is the relation between $\beta_{c}$ and $\boldsymbol{g}(\beta)$ in $\mathcal{C}$ and $\mathcal{A}$ as $N_{f} \rightarrow \infty$ in the limit? 

ii) What is the behaviour of quantity \eqref{eq:F} at $\beta_{c}$ in $\mathcal{C}$ and $\mathcal{A}$ for intermediate (non-limiting) values of $P_{g}$ as $N_{f} \rightarrow \infty$ in the limit?
\\
\\
To calculate the value of $\beta_{c}$ for $\boldsymbol{g}$ we proceed as follows. For both $\mathcal{C}$ and $\mathcal{A}$ when $P_{g} \rightarrow 0$ in the limit, following any growth event an absorbing state will always be reached before the next growth event occurs. This means $\langle F^{\boldsymbol{g}(\beta)} \rangle_{P_{g} \rightarrow 0}$ can considered as a series of individual absorption events:
\begin{align}
\langle F^{\boldsymbol{g}(\beta)} \rangle_{P_{g} \rightarrow 0} = \sum_{N=N_{s}-1}^{N_{f}-1}\boldsymbol{g}_{N}\boldsymbol{f}_{N},
\label{eq:contproof2}
\end{align}
where $\boldsymbol{f}_{N}$ is a column vector and $\boldsymbol{f}_{N}(k)$ is the expected number of flips before an absorbing state is reached given the new site is site $k$. We have included the term $\boldsymbol{g}_{N_{s}-1}\boldsymbol{f}_{N_{s}-1}$ to represent the expected number of flips before the absorption state is reached from the initial distribution of states, and define $\boldsymbol{g}_{N_{s}-1} \equiv 1$. Expanding the RHS of Eq. \eqref{eq:contproof2} in the following manner
\begin{align}
\langle F^{\boldsymbol{g}(\beta)} \rangle_{P_{g} \rightarrow 0} = & \ \beta\left(\boldsymbol{f}_{N_{S}-1} + \sum_{N = N_{s}}^{N_{f}-1}\left(\sum_{k=2}^{N} \left(\frac{\boldsymbol{f}_{N}(k)}{N-1}\right)\right) \right) \nonumber \\ &+ (1-\beta)\left(\boldsymbol{f}_{N_{S}-1} + \sum_{N=N_{s}}^{N_{f}-1} \boldsymbol{f}_{N}(1)\right),
\label{eq:contproof3}
\end{align}
and rewriting Eq. \eqref{eq:contproof3} as
\begin{align}
\langle F^{\boldsymbol{g}(\beta)} \rangle_{P_{g} \rightarrow 0} = \beta\langle F^{\boldsymbol{g}(\beta = 1)} \rangle_{P_{g} \rightarrow 0} + (1-\beta)\langle F^{\boldsymbol{g}(\beta=0)} \rangle_{P_{g} \rightarrow 0},
\label{eq:contproof}
\end{align}
we then impose Eq. \eqref{eq:a2} to obtain
\begin{align}
\beta_{c} = \frac{\langle F^{\boldsymbol{g}(\beta)} \rangle_{P_{g} \rightarrow \infty} - \langle F^{\boldsymbol{g}(\beta = 0)} \rangle_{P_{g} \rightarrow 0}}{\langle F^{\boldsymbol{g}(\beta = 1)} \rangle_{P_{g} \rightarrow 0} - \langle F^{\boldsymbol{g}(\beta = 0)} \rangle_{P_{g} \rightarrow 0}}.
\label{eq:betac}
\end{align}
In deriving Eq. \eqref{eq:betac} we have assumed that
\begin{align}
\langle F^{\boldsymbol{g}(\beta = 1)} \rangle_{P_{g} \rightarrow 0} \geq \langle F^{\boldsymbol{g}(\beta)} \rangle_{P_{g} \rightarrow \infty} \geq \langle F^{\boldsymbol{g}(\beta = 0)} \rangle_{P_{g} \rightarrow 0}, \ \forall \ N_{f},
\label{eq:betac_assum}
\end{align}
for both $\mathcal{C}$ and $\mathcal{A}$. 

Using Eq. \eqref{eq:betac} we plot $\beta_{c}$ as a function of $N_{f}$ for both $\mathcal{C}$ and $\mathcal{A}$ in Fig. \ref{fig:figure2} (c)\footnote{$\beta_{c}$ can be computed efficiently by using Eq. \eqref{eq:all_two} in tandem with the observation that in the limit $P_{g} \rightarrow 0$, following any growth event $\mathcal{C}$ can only be in one of $2N$ states. From these $2N$ states $\mathcal{C}$ can only reach $N(N+1)$ states, including the absorbing states. Therefore, $K$ can be calculated using a matrix of $O(N^{2})$ instead of $O(2^{N})$. Similar reasoning applies to computing $\beta_{c}$ for $\mathcal{A}$.}. At $N_{f} = 3$, $\beta_{c} = 0$ for both $\mathcal{C}$ and $\mathcal{A}$. In the (thermodynamic) limit, $N_{f} \rightarrow \infty$, it appears $\beta_{c} \rightarrow 1$ for $\mathcal{C}$ and $\beta_{c} \rightarrow 0$ for $\mathcal{A}$. In the case of $\mathcal{C}$, $\beta_{c}$ grows monotonically from 0 to 1.  Whereas for $\mathcal{A}$, $\beta_{c}$ first rises from 0 till $N_{f} = 7$ before beginning to decrease again. It is also important to stress that the value $\beta_{c}$ takes is independent of the functional form of the growth, $N(t)$. For instance, linear or logistic growth would also result in the same value of $\beta_{c}$ in the limit $P_{g} \rightarrow 0$.
\\
\\
\indent To better characterise the dependence of $\beta_{c}$ on $N_{f}$ in $\mathcal{C}$ we study how the expected number of flips evolves in the limit $P_{g} \rightarrow 0$. We begin with a lattice of size $N_{s} = 2$ and $\rho_{0} = 0.5$, and so the four initial states are
$$(0,0), (1,0), (0,1), (1,1),$$
which all occur with equal frequency. States $(0,0)$ and $(1,1)$ are already in an absorbing state for $\mathcal{C}$ and hence no flips occur.  States $(1,0)$ and $(0,1)$ each only require 1 flip to reach an absorbing state.  Therefore, from these four states the expected number of flips before an absorbing state is reached is 0.5. Following a growth event, using $\boldsymbol{g}_{2}$, the possible states are
$$(\bar{0},0,0), (\bar{1},0,0), (0,\bar{0},0), (0,\bar{1},0),
(\bar{0},1,1), (\bar{1},1,1), (1,\bar{0},1), (1,\bar{1},1),$$
where the bar indicates the location of the new site. 
With some probability dependent on $\rho_{0}$, the new state will already be in an absorbing state.  If not, it will be in a state of the form
$$(\bar{1},0,...), (0,\bar{1},0,...), (0,0,\bar{1},0,...),(0,0,0,\bar{1},0,...), ... \ ,$$
or
$$(\bar{0},1,...), (1,\bar{0},1,...), (1,1,\bar{0},1,...),(1,1,1,\bar{0},1,...), ... \ .$$
To calculate the expected number of flips till an absorbing state in reached we sum the expected number of flips associated with these states multiplied by their probability of occurrence, which depends on the values of $\beta$ and $\rho_{0}$. For example, when $\beta = 0$, $N_{s} = 2$ and $\rho_{0} = 0.5$ we begin with $(0,0)$, $(0,1)$, $(1,0)$, $(1,1)$, from which we can grow to
$$(\bar{0},0,0), (\bar{1},0,0),
(\bar{0},1,1), (\bar{1},1,1),$$
and from the following growth event we can grow into
$$(\bar{0},0,0,0), (\bar{1},0,0,0),
(\bar{0},1,1,1), (\bar{1},1,1,1),$$
and so on. In this instance the expected number of flips from $N_{s} = 2$ to $N_{f}$ has the simple form
\begin{align}
\langle F_{\mathcal{C}}^{\boldsymbol{g}(\beta = 0), \rho_{0}} \rangle_{P_{g} \rightarrow 0} = \rho_{0}(1-\rho_{0})\left(\frac{(N_{f}(N_{f}+1))}{2} - 1 \right).
\label{eq:rho05}
\end{align}
In Eq. \eqref{eq:rho05} the value of $\rho_{0}$ only affects the prefactor of the leading order term, $O(N_{f}^{2})$, and not the order of the leading order term.
Equation \eqref{eq:rho05} is also the case for $\mathcal{A}$ when $\rho_{0} = 0.5$.
The expected number of flips from $N_{s} = 2$ to $N_{f}$ when $\beta = 1$ in $\mathcal{C}$ in the limit $P_{g} \rightarrow 0$ is
\begin{align}
\langle F_{\mathcal{C}}^{\boldsymbol{g}(\beta = 1), \rho_{0}} \rangle_{P_{g} \rightarrow 0} = 4\rho_{0}(1-\rho_{0})\langle F_{\mathcal{C}}^{\boldsymbol{g}(\beta = 1), \rho_{0}=0.5} \rangle_{P_{g} \rightarrow 0},
\end{align}
and so again changing $\rho_{0}$ only affects the prefactor of the leading order term.

More generally for $\mathcal{C}$ as $P_{g} \rightarrow 0$ in the limit the leading terms evolve as
\begin{align}
\boldsymbol{c}_{N} = \left(\frac{N}{2}, N, N, N,...\right),
\label{eq:c_vector1}
\end{align}
where $\boldsymbol{c}_{N}(1)$ is the leading order term of the expected number of flips till an absorbing state is reached from state $(\bar 1,0,0,0,0,...)$ or $(\bar{0},1,1,1,1,...)$ on a lattice of size $N$, $\boldsymbol{c}_{N}(2)$ is the leading order term of the expected number of flips till an absorbing state is reached from state $(0,\bar{1},0,0,0,...)$ or $(1,\bar{0},1,1,1,...)$ on a lattice of size $N$, and so on. 
We now sum the terms in Eq. \eqref{eq:c_vector1} to $N_{f}$
\begin{align}
\boldsymbol{\tilde c}_{N_{f}} = \left(\frac{N_{f}^2}{4}, \frac{N_{f}^2}{2}, \frac{N_{f}^2}{2},\frac{N_{f}^2}{2},...\right),
\label{eq:c_vector2}
\end{align}
where $\boldsymbol{\tilde c}_{N_{f}}(1)$ is the leading order term in the limit $P_{g} \rightarrow 0$ when the new site is always site 1, $\boldsymbol{\tilde c}_{N_{f}}(2)$ is the leading order term in the limit $P_{g} \rightarrow 0$ when the new site is always site 2, and so on. From Eq. \eqref{eq:c_vector2} it is evident that any combination of internal growth events in the limit $P_{g} \rightarrow 0$ evolves equivalently in its leading order term.
Setting $\rho_{0} = 0.5$ gives us
\begin{align}
\boldsymbol{\tilde c}_{N_{f}} = \left(\frac{N_{f}^2}{8}, \frac{N_{f}^2}{4}, \frac{N_{f}^2}{4},\frac{N_{f}^2}{4},...\right).
\label{eq:c_vector3}
\end{align}

To obtain $\langle F_{\mathcal{C}}^{\boldsymbol{g}(\beta),\rho_{0} = 0.5}\rangle_{P_{g}\rightarrow \infty}$ we use the following numerically observed identity:
\begin{align}
\langle F_{\mathcal{C}}^{\boldsymbol{g}(\beta),\rho_{0} = 0.5}\rangle_{P_{g}\rightarrow \infty} = \left(\frac{1}{4\binom{N_{f}-2}{j-1}}\right)\sum_{q=1}^{\binom{N_{f}}{j}}\boldsymbol{k}(q), \ j < N_{f}/2, \ N_{f} > 2,
\label{eq:all_two}
\end{align}
where $j$ indicates how many sites are in state `1', $q$ indexes the states containing $j$ `1s' on a lattice of size $N_{f}$, and $\boldsymbol{k}(q)$ is the expected number of flips till an absorbing state is reached from state $q$. For instance, if $N_{f} = 5$ and $j = 1$, $\boldsymbol{k}(1)$ is the exact number of flips to an absorbing state starting from state $(1,0,0,0,0)$, $\boldsymbol{k}(2)$ is the exact number of flips to an absorbing state starting from state $(0,1,0,0,0)$, and so on.
In the case of $j = 1$ this is simply:
\begin{align}
\langle F^{\boldsymbol{g}(\beta),\rho_{0}=0.5}\rangle_{P_{g}\rightarrow \infty} = \left(\frac{1}{4}\right)\sum_{i=1}^{N_{f}}\boldsymbol{k}(i) \sim \left(\frac{1}{4}\right)\sum_{i=1}^{N_{f}}\boldsymbol{c}_{N_{f}}(i) \sim \left(\frac{1}{4} \right)N_{f}^{2},
\label{eq:inf}
\end{align}
where we have used Eq. \eqref{eq:c_vector1}. The effect of $\rho_{0}$ on $\langle F_{\mathcal{C}}^{\boldsymbol{g}(\beta),\rho_{0}}\rangle_{P_{g}\rightarrow \infty}$ is
$$\langle F_{\mathcal{C}}^{\boldsymbol{g}(\beta),\rho_{0}}\rangle_{P_{g}\rightarrow \infty} = 4\rho_{0}(1-\rho_{0})\langle F_{\mathcal{C}}^{\boldsymbol{g}(\beta),\rho_{0} = 0.5}\rangle_{P_{g}\rightarrow \infty}.$$
Whereas the effect of $\rho_{0}$ on $\langle F_{\mathcal{A}}^{\boldsymbol{g}(\beta),\rho_{0}}\rangle_{P_{g}\rightarrow \infty}$ is
$$\langle F_{\mathcal{A}}^{\boldsymbol{g}(\beta),\rho_{0}}\rangle_{P_{g}\rightarrow \infty} = \langle F_{\mathcal{A}}^{\boldsymbol{g}(\beta),\rho_{0}=0.5}\rangle_{P_{g}\rightarrow \infty} + \left(\frac{1 - 4\rho(1-\rho)}{4}\right)N_{f}.$$
We now analyze $\mathcal{A}$ as $P_{g} \rightarrow 0$ in the limit. As before we begin with a lattice of size $N_{s} = 2$ and with $\rho_{0} = 0.5$, and so this means the four possible initial states
$$(0,0), (1,0), (0,1), (1,1),$$
occur in equal frequency. States $(1,0)$ and $(0,1)$ are already in an absorbing state hence no flips occur.  States $(0,0)$ and $(1,1)$ each only require 1 flip to reach an absorbing state.  Therefore, starting from these four states in equal frequency the expected number of flips before an absorbing state is reached is 0.5. Following a growth event, using $\boldsymbol{g}_{2}$, the possible states for $\mathcal{A}$ are
$$(\bar{0},1,0), (\bar{1},1,0), (1,\bar{0},0), (1,\bar{1},0),
(\bar{0},0,1), (\bar{1},0,1), (0,\bar{0},1), (0,\bar{1},1).$$
In $\mathcal{A}$ only a growth event in which the new site is site $i = 1$ can result in an absorbing state, and no internal growth event can result in an absorbing state. Furthermore, for an internal growth event the amount of sites that have to flip before an absorbing state is reached is at least the shortest distance of the new site from one of the two boundaries of the lattice. This demonstrates a key difference between $\mathcal{C}$ and $\mathcal{A}$, which can be considered as a type of boundary effect, and is a recurrent theme in the physics of growing active matter that we return to in the discussion.

Following an internal growth event, in the limit $P_{g} \rightarrow 0$, $\mathcal{A}$ will either be in the state
$$(1,\bar{1},0,1,0,...), (0,\bar{1},1,0,1,...),(1,0,\bar{1},1,0,...),(0,1,\bar{1},0,1,...), ... ,$$
or
$$(1,\bar{0},0,1,0,...), (0,\bar{0},1,0,1,...), (1,0,\bar{0},1,0,...),(0,1,\bar{0},0,1,...), ...$$
and so in $\mathcal{A}$ growth events at the same position can result in two different non-absorbing states.  This means the expected number of flips for growth events at a single position has to be averaged over both possibilities at the correct frequencies.

In $\mathcal{A}$ as $P_{g} \rightarrow 0$ in the limit the leading terms evolve as
\begin{align}
\boldsymbol{a}_{N} = \left(\frac{N}{2}, N, 2N, 3N, 4N, 5N, 6N,...\right).
\label{eq:a_vector1}
\end{align}
If we set $\rho_{0} = 0.5$ and sum Eq. \eqref{eq:a_vector1} to $N_{f}$ we obtain
\begin{align}
\boldsymbol{\tilde a}_{N_{f}} = \left(\frac{N_{f}^2}{8},\frac{N_{f}^2}{2},N^2,\frac{3N_{f}^2}{2}, 2N_{f}^2, \frac{5N_{f}^2}{2}, 3N_{f}^2,...\right).
\label{eq:a_vector3}
\end{align}
Equation \eqref{eq:a_vector3} demonstrates that for $\mathcal{A}$ with $\boldsymbol{g}_{N}$ and $\beta = 1$ the leading order terms are $O(N_{f}^3)$. The exact value for $\mathcal{A}$ when $\rho_{0} = 0.5$ and $\beta = 1$ for $\hat{g}$ is
\begin{align}
\langle F_{\mathcal{C}}^{\boldsymbol{g}(\beta=1),\rho_{0}=0.5}\rangle_{P_{g}\rightarrow 0} = \frac{1}{18}\left(N_{f}^3 + 8N_{f} - 15\right),
\label{eq:a_vector_rho05}
\end{align}
and more generally 
$$\langle F_{\mathcal{C}}^{\boldsymbol{g}(\beta=1),\rho_{0}}\rangle_{P_{g}\rightarrow 0} = O(N_{f}^3), \ \forall \ \rho_{0}.$$

We now return to Eq. \eqref{eq:betac}. Using Eqs. \eqref{eq:rho05}, \eqref{eq:c_vector3} and  \eqref{eq:inf} in the case of $\mathcal{C}$ for $\boldsymbol{g}$ we have
\begin{align}
\lim_{N_{f}\rightarrow \infty}\beta_{c} \sim \lim_{N_{f}\rightarrow \infty}\left(\frac{\frac{1}{4}N_{f}^2 - \frac{1}{8}N_{f}^2}{\frac{1}{4}N_{f}^2 - \frac{1}{8}N_{f}^2}\right) \rightarrow 1,
\label{eq:g_vector1_solC}
\end{align}
and using Eqs. \eqref{eq:rho05}, \eqref{eq:inf} and \eqref{eq:a_vector_rho05} in the case of $\mathcal{A}$ for $\boldsymbol{g}$ we have
\begin{align}
\lim_{N_{f}\rightarrow \infty}\beta_{c} \sim \lim_{N_{f}\rightarrow \infty}\left(\frac{\frac{1}{4}N_{f}^2 - \frac{1}{8}N_{f}^2}{\frac{1}{18}N_{f}^3 - \frac{1}{8}N_{f}^2}\right) \rightarrow 0.
\label{eq:g_vector1_solA}
\end{align}

To corroborate expressions \eqref{eq:g_vector1_solC} and \eqref{eq:g_vector1_solA}, which have not been formally proven, we examine the following two growth vectors\footnote{We now refer to the original $\boldsymbol{g}$ studied as $\boldsymbol{g}^{1}$.}:
\begin{align}
\boldsymbol{g}^{2}_{N} = (1-\beta,\beta, 0, ...,0),
\label{eq:g_vector2}
\end{align}
and
\begin{align}
\boldsymbol{g}^{3}_{N} = (1-\beta,\beta/2, \beta/2, 0, ...,0).
\label{eq:g_vector3}
\end{align}
In $\boldsymbol{g}^{2}$ and $\boldsymbol{g}^{3}$ the interpretation of $\beta$ is as before, whereby $\beta = 0$ represents the most conservative implementation of growth, and $\beta = 1$ is the most disruptive implementation of growth. When $N = 2$, $\boldsymbol{g}^{3}$ is not well-defined, and so we use $\boldsymbol{g}^{2}$.
We also examine
\begin{align}
\boldsymbol{g}^{4}_{N} = (1-\beta, \beta^{2}, \beta^{3}, ...,\beta^{N})/\sum_{i}\boldsymbol{g}^{4}_{N}(i),
\label{eq:g_vector4}
\end{align}
and $\boldsymbol{g}^{4}$ is equivalent to $\boldsymbol{g}^{1}$ when $\beta = 0$ or $\beta = 1$. 

\begin{figure}[H]
\centering
    \begin{subfigure}{.4\textwidth}
	\includegraphics[width=\textwidth]{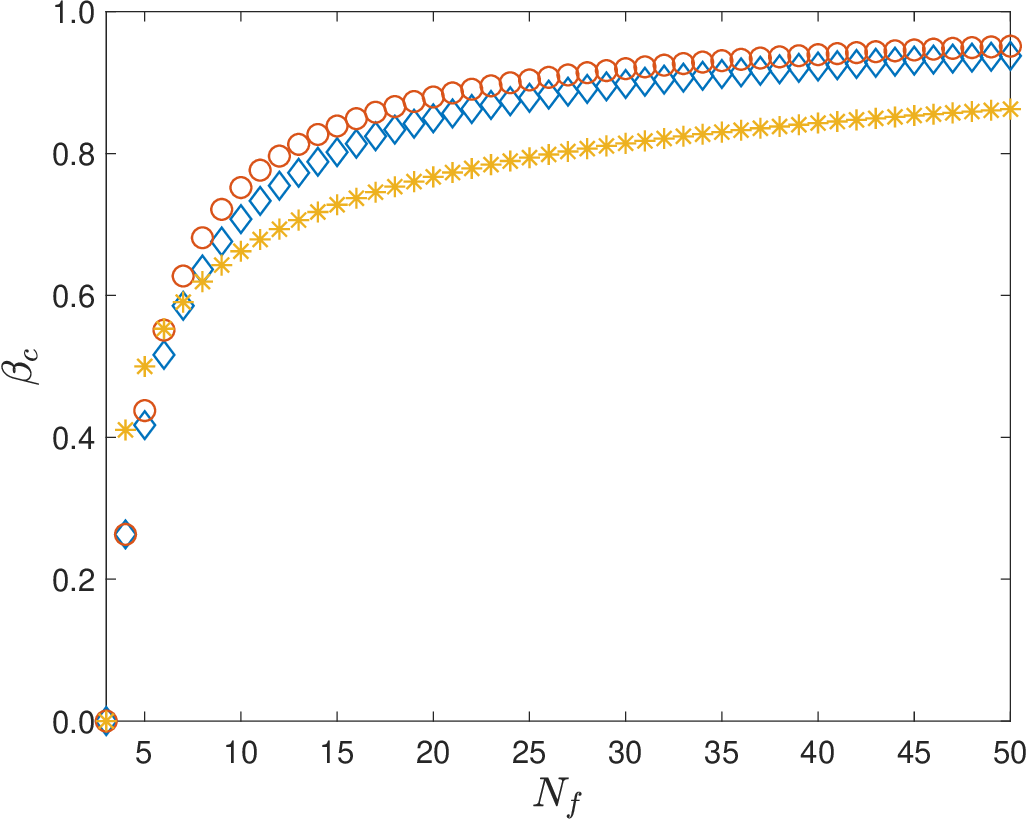}
	\subcaption{}
	\end{subfigure}
	~
	\begin{subfigure}{.4\textwidth}
	\includegraphics[width=\textwidth]{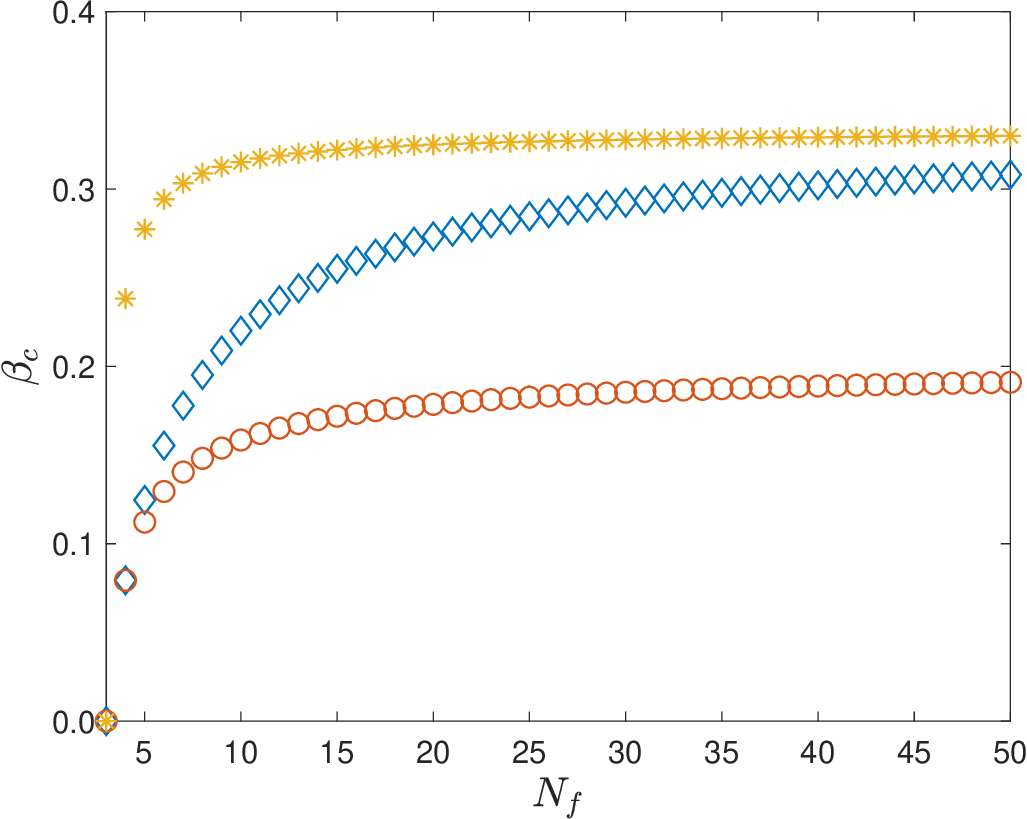}
	\subcaption{}
	\end{subfigure}
\caption{The value of $\beta_{c}$ in systems grown from $N_{s} = 2$ to $N_{f}$, with $P_{d} = 1$ and $\rho_{0} = 0.5$. (a) $\mathcal{C}$, (b) $\mathcal{A}$, and $\boldsymbol{g}^{2}_{N}$ (blue diamonds), $\boldsymbol{g}^{3}_{N}$ (red circle), and $\boldsymbol{g}^{4}_{N}$ (yellow stars). Data points were generated using Eq. \eqref{eq:F}.}
\label{fig:figure3}
\end{figure}

In Fig. \ref{fig:figure3} (a) we plot the evolution of $\beta_{c}$ in $\mathcal{C}$ for $\boldsymbol{g}^{2}$, $\boldsymbol{g}^{3}$ and $\boldsymbol{g}^{4}$. For $\mathcal{C}$ $\beta_{c} \rightarrow 1$ for all $\boldsymbol{g}$. This is because all internal growth events evolve $(1/4)N_{f}^{2}$, which is the same as $\langle F^{\boldsymbol{g}(\beta)} \rangle_{P_{g} \rightarrow \infty}$.
For $\mathcal{A}$ this is not the case. In $\mathcal{A}$ with $\boldsymbol{g}^{2}$ for $\beta_{c}$ we have
\begin{align}
\lim_{N_{f}\rightarrow \infty}\beta_{c} \sim \lim_{N_{f}\rightarrow \infty}\left(\frac{\frac{1}{4}N_{f}^2 - \frac{1}{8}N_{f}^2}{\frac{1}{2}N_{f}^2 - \frac{1}{8}N_{f}^2}\right) \rightarrow \frac{1}{3},
\label{eq:g_vector2_sol}
\end{align}
and $\beta_{c}$ for $\boldsymbol{g}^{3}$ is
\begin{align}
\lim_{N_{f}\rightarrow \infty}\beta_{c} \sim \lim_{N_{f}\rightarrow \infty}\left(\frac{\frac{1}{4}N_{f}^2 - \frac{1}{8}N_{f}^2}{\frac{3}{4}N_{f}^2 - \frac{1}{8}N_{f}^2}\right) \rightarrow \frac{1}{5}.
\label{eq:g_vector3_sol}
\end{align}
In the case of $\boldsymbol{g}^{4}$ isolating $\beta_{c}$ as in Eq. \eqref{eq:betac} is not possible. However, asymptotically the following must hold when $\beta$ is $\beta_{c}$
\begin{align}
\lim_{N_{f} \rightarrow \infty}\left(\frac{1}{4}N_{f}^{2} \sim \hat\beta^{-1}\sum^{\infty}_{j=2}\beta^{j}\left(\frac{j-1}{2}\right)N_{f}^{2} + \hat\beta^{-1}(1 - \beta)\frac{1}{8}N_{f}^{2}\right),
\label{eq:g_vector4_asym1}
\end{align}
where $\hat\beta$ is the asymptotic normalization factor
\begin{align}
\hat\beta = (1 - \beta_{c}) + \sum^{\infty}_{j=2}\beta_{c}^{j}.
\label{eq:g_vector4_asym2}
\end{align}
Expression \eqref{eq:g_vector4_asym1} holds when $\beta_{c} = 1/3$. The data points in Fig. \ref{fig:figure3} (b), calculated numerically using Eq. \eqref{eq:F}, validate expressions \eqref{eq:g_vector2_sol}-\eqref{eq:g_vector4_asym1} and corroborate expressions \eqref{eq:g_vector1_solC} and \eqref{eq:g_vector1_solA}, which were derived analytically using assumption based on numerical observations.

\subsubsection{Intermediate values of $P_{g}$ and the role of $N(t)$}
\label{sect:PgNt}

In the previous section we used Eq. \eqref{eq:all_two} to relate the expected number of flips as $P_{g} \rightarrow 0$ in the limit to the expected number of flips as $P_{g} \rightarrow \infty$ in the limit. The limit $P_{g} \rightarrow 0$ can be thought of as a quasistatic transition, whereby $\mathcal{C}$ or $\mathcal{A}$ are always allowed to find an absorbing state before the next growth event occurs, whereas the limit $P_{g} \rightarrow \infty$ can be thought of as initializing $\mathcal{C}$ or $\mathcal{A}$ at $N_{f}$. This means that in both of these limiting regimes the more complex interactions between growth and self-organization in determining the expected number of flips in $\mathcal{C}$ or $\mathcal{A}$ are not present, and we could also ignore the nature of $N(t)$. For $\mathcal{C}$ or $\mathcal{A}$ grown with intermediate values of $P_{g}$, which we indicate by $\bar P_{g}$, this is not the case.

To better characterize the interaction between $\bar P_{g}$ and $N(t)$ in the evolution of $\mathcal{C}$ and $\mathcal{A}$ we decompose $\langle F^{\boldsymbol{g}(\beta_{c})} \rangle_{\tilde P_{g}}$ in the following manner:
\begin{align}
\langle F^{\boldsymbol{g}(\beta_{c})} \rangle_{\tilde P_{g}} = f^{N_{s} \rightarrow N_{f}} + f^{N_{f}},
\label{eq:pginvar_weak2}
\end{align}
where $f^{N_{s} \rightarrow N_{f}}$ is the expected number of flips as the lattice grows from $N_{s}$ to $N_{f}$, but not including flips when the lattice is at size $N_{f}$, and $f^{N_{f}}$ is the expected number of flips while the lattice is size $N_{f}$ until an absorbing state is reached. It is clear that when $P_{g} \rightarrow \infty$ in the limit
$$f^{N_{s} \rightarrow N_{f}} = 0,$$
and
$$f^{N_{f}} = \langle F^{\boldsymbol{g}(\beta)} \rangle_{P_{g} \rightarrow \infty}.$$
Similarly, when $P_{g} \rightarrow 0$ in the limit
$$f^{N_{s} \rightarrow N_{f}} = \langle F^{\boldsymbol{g}(\beta_{c})} \rangle_{P_{g} \rightarrow 0} - f^{N_{f}}.$$
The first possibility we consider is
\begin{align}
\lim_{N_{f} \rightarrow \infty}\frac{f^{N_{s}\rightarrow N_{f}}}{f^{N_{f}}} \rightarrow 0,
\label{eq:pginvar_weak3}
\end{align}
whereby if $O(f^{N_{f}}) \rightarrow O(\langle F^{\boldsymbol{g}(\beta)} \rangle_{P_{g \rightarrow \infty}})$ as $N_{f} \rightarrow \infty$ in the limit this combination of $\bar P_{g}$ and $N(t)$ asymptotically behave like $P_{g} \rightarrow \infty$ in the limit. Intuitively, this can be understood as the lattice is growing faster than the voter model can self-organize. 
As an example let us consider the exponential growth we have examined in this work, in which
\begin{align}
O(f^{N_{s} \rightarrow N_{f}}) \sim \sum_{N = N_{s}}^{N_{f}-1}\epsilon N.
\label{eq:pginvar_weak7}
\end{align}
In exponential growth the ratio of the expected number of growth events with the expected number of attempted decision events in a given interval is constant for both $\mathcal{C}$ and $\mathcal{A}$, i.e. $P_{d}/P_{g}$, and so at most, $\epsilon \sim O(\frac{1}{N})$. This means for exponential growth we have
\begin{align}
O(\langle F^{\boldsymbol{g}(\beta_{c})} \rangle_{\tilde P_{g}}) \sim & \ O(N_{f}) + O(f^{N_{f}}), \nonumber \\
\sim & \ O(f^{N_{f}}),
\label{eq:pginvar_weak8}
\end{align}
as $O(f^{N_{f}}) \sim N_{f}^{2}$, and $\bar P_{g}$ and exponential growth interact to behave like $P_{g} \rightarrow \infty$ in the limit so long as $O(f^{N_{f}}) \rightarrow \frac{1}{4}N_{f}^{2}$ as $N_{f} \rightarrow \infty$ in the limit\footnote{An argument of this nature holds for any functional form of the growth greater than linear, so that $\epsilon < O(1)$.}. 

For $\mathcal{C}$ numerical results suggest
\begin{align}
\langle F_{\mathcal{C}}^{\boldsymbol{g}(\beta = 1)} \rangle_{P_{g} \rightarrow 0} \geq f^{N_{s} \rightarrow N_{f}} + f^{N_{f}} \geq \langle F_{\mathcal{C}}^{\boldsymbol{g}(\beta_{c})} \rangle_{P_{g} \rightarrow \infty}, \ \forall \ N_{f},
\label{eq:betac_assum2}
\end{align}
and so in the case of exponential growth $O(f^{N_{f}}) \rightarrow \frac{1}{4}N_{f}^{2}$ for all $\boldsymbol{g}$ and $P_{g}$. We refer to this behavior as $P_{g}$-invariance for some $N(t)$. $P_{g}$-invariance can be thought of as a process becoming \emph{conservative}: by which we mean at $\beta_{c}$ the leading order term in the expected number of flips in $\mathcal{C}$ or $\mathcal{A}$ becomes independent of re-scalings of the path $N(t)$, that is, changing the value of $P_{g}$. More generally, expression \eqref{eq:betac_assum2} appears to hold for all $N(t)$ in the case of $\mathcal{C}$, for instance with linear growth. This more general property we refer to as $N(t)$-invariance, whereby at $\beta_{c}$ for a given $\boldsymbol{g}$, the leading order term for the expected number of flips in $\mathcal{C}$ or $\mathcal{A}$ becomes independent of any path taken from $N_{s}$ to $N_{f}$, so long as $N(t)$ is monotonically increasing.  
A similar assumption as expression \eqref{eq:betac_assum2} cannot be employed for $\mathcal{A}$ however, one reason being that $\langle F_{\mathcal{A}}^{\boldsymbol{g}(\beta = 0)} \rangle_{P_{g} \rightarrow 0} \sim \frac{1}{8}N_{f}^{2}$. 

Conversely, one can also imagine a functional form of growth whereby
\begin{align}
\lim_{N_{f} \rightarrow \infty}\frac{f^{N_{s} \rightarrow N_{f}}}{f^{N_{f}}} \rightarrow \infty,
\label{eq:pginvar_weak4}
\end{align}
such that $f^{N_{s} \rightarrow N_{f}} \rightarrow \langle F^{\boldsymbol{g}(\beta_{c})} \rangle_{P_{g} \rightarrow 0} - f^{N_{f}}$. In this instance this combination of $\bar P_{g}$ and $N(t)$ would asymptotically behave like $P_{g} \rightarrow 0$ in the limit. Expressions \eqref{eq:pginvar_weak3} and \eqref{eq:pginvar_weak4} can be envisaged as postulating `basins of attraction' for models $\mathcal{C}$ and $\mathcal{A}$, whereby irrespective of the value of $\bar P_{g}$ the expected number of flips asymptotically approaches the value associated with either $P_{g} \rightarrow 0$ or $P_{g} \rightarrow \infty$ in the limit. Behavior such as this would be reminiscent of renormalization group methods \cite{Goldenfeld2018}.
The final possibility to account for is
\begin{align}
\lim_{N_{f} \rightarrow \infty}\frac{f^{N_{s} \rightarrow N_{f}}}{f^{N_{f}}} \rightarrow \kappa,
\end{align}
where $\kappa$ is a constant greater than zero. In this instance the growth rate, $\mathrm{d}N(t)/\mathrm{d}t$, is decreasing in proportion to the increasing complexity of finding an absorbing state as $N_{f} \rightarrow \infty$ in the limit.

\section{Discussion}

As suggested in the main text, the evolution of $\beta_{c}$ in models $\mathcal{C}$ and $\mathcal{A}$ is perhaps best understood as a boundary effect. In the case of $\mathcal{C}$ as $P_{g} \rightarrow 0$ in the limit all internal growth events result in the same leading order behaviour for the expected number of flips before an absorbing state is achieved, and only the leading order term at the boundary site evolves differently.  Conversely, in $\mathcal{A}$ as $P_{g} \rightarrow 0$ in the limit the leading order behaviour for each site is determined by its distance from the nearest boundary. It is these boundary-related effects that are being balanced when $\beta_{c}$ is being calculated for a given $\boldsymbol{\hat{g}}$ in $\mathcal{C}$ or $\mathcal{A}$. The role of boundary effects in determining the behavior of growing active matter has been reported in other modeling approaches. For instance, it has been shown in population models on growing lattices that the implementation of growth can determine the competition outcome by manipulating the evolution of spatial correlations \cite{Ross2016b,Ross2017b}. In this study, one of the two growth implementations studied requires an `origin of growth', i.e. a boundary, to be specified.  This origin of growth breaks both translational and scale invariances present in the other growth implementation studied, and so explains the different competition outcomes for these two growth implementations. 
A boundary effect has also been shown to be responsible for the generation of a traveling wave in a network growth model \cite{Ross2018c}. In this model network growth is coupled to the dynamics of a random walker situated on the network, and the boundary effect can be envisaged as the boundary of the network `chasing' the walker, as such random walker travels like a wave in the age-space of the network.

Another useful framework for understanding the physics of growing active matter is as the manipulation of relaxation times. For example, in the work presented here as $P_{g} \rightarrow 0$ in the limit the relaxation time to find the next absorbing state following a growth event is a function of the distance from the boundary in the case of $\mathcal{A}$. For $\mathcal{C}$ this is not the case, as all internal growth events result in the same leading order behavior before the next absorbing state is found. This difference in relaxation times following a growth event in the limit $P_{g} \rightarrow 0$ is what underpins the differences between $\mathcal{C}$ and $\mathcal{A}$. Returning to the network growth model previously mentioned, the traveling wave is generated by repeatedly sampling the position of random walker as it forever tries (and fails) to equilibrate on an ever-growing network \cite{Ross2018b,Ross2018c}. These examples demonstrate how phenomena observed in the physics of growing active matter are often due to the growth of a space interacting with a process whose equilibration/relaxation/fixation/information propagation `rate' is held constant in some way.   

Finally, it is interesting to consider what other macroscopic properties beyond the expected number of flips could be controlled by the implementation of growth. For instance, different energetic costs could be associated with different state transitions, and in this way the implementation of growth could be used to control the energetic cost of self-organisation. However, in this case we may want to associate an energetic cost with decision events whether they result in a flip or not, the expected number of which tend to infinity as $P_{g} \rightarrow 0$ in the limit.

\section*{Conflict of interest}

The authors declare that they have no conflict of interest.

\end{document}